\documentclass[12pt,preprint]{article}
\usepackage{amssymb}
\usepackage{amsmath}
\usepackage{graphics}
\usepackage{epsfig}

\newcommand{\eqb}{\begin{equation}}
\newcommand{\eqe}{\end{equation}}
\newcommand{\dmb}{\begin{displaymath}}
\newcommand{\dme}{\end{displaymath}}

\newcommand{\eab}{\begin{eqnarray}}
\newcommand{\eae}{\end{eqnarray}}

\newcommand{\be}{\begin{equation}}
\newcommand{\ee}{\end{equation}}

\begin{document}

\begin{titlepage}
\begin{flushright} 
KA-TP-07-2007
\end{flushright}

\vspace{0.6cm}

\begin{center}
\Large{Linear growth of the trace anomaly in 
Yang-Mills thermodynamics}
\vspace{1.5cm}

\large{Francesco Giacosa$^\dagger$ and Ralf Hofmann$^*$}

\end{center}
\vspace{1.5cm} 

\begin{center}
{\em $\mbox{}^\dagger$ Institut f\"ur Theoretische Physik\\ 
Universit\"at Frankfurt\\ 
Johann Wolfgang Goethe - Universit\"at\\ 
Max von Laue--Str. 1\\ 
60438 Frankfurt, Germany}
\end{center}
\vspace{1.5cm}

\begin{center}
{\em $\mbox{}^*$ Institut f\"ur Theoretische Physik\\ 
Universit\"at Karlsruhe (TH)\\ 
Kaiserstr. 12\\ 
76131 Karlsruhe, Germany}
\end{center}
\vspace{1.5cm}

\begin{abstract}

In the lattice work by Miller \cite{millerrev,miller97} and 
in the work by Zwanziger \cite{zw} a linear growth of 
the trace anomaly for high temperatures was found in pure 
SU(2) and SU(3) Yang-Mills theories. These results show the remarkable property  that 
the corresponding systems are strong interacting even at high temperatures.
We show that within an analytical 
approach to Yang-Mills thermodynamics this 
linear rise is obtained and is directly connected to the presence of a temperature-dependent ground state,
which describes (part of) the nonperturbative nature of the Yang-Mills system.
Our predictions are in approximate agreement with \cite{millerrev,miller97,zw}.  
        
\end{abstract} 

\end{titlepage}\bigskip

\section{Introduction}

Studies of pure SU(2) and SU(3) Yang-Mills theories at finite temperature $T$
represent an important subject in modern high energy physics and a
fundamental step toward the analysis of the properties of the
Quark-Gluon-Plasma. In particular, the question if and for which
temperatures a gas of (almost) free gluons offers a valid approximate
description of such strong interacting theories goes to the heart of the
problem. A key-quantity to address this issue is the trace of the stress
energy tensor $\theta _{\mu \mu }\equiv \rho -3p,$ where $\rho $ and $p$ are
the energy density and the pressure of the system. In fact, the trace $%
\theta _{\mu \mu }$ vanishes for a free gas of massless gluons, being a
consequence of conformal invariance of Yang-Mills theories. Thus, one would
expect that $\theta _{\mu \mu }$ vanishes in the high-temeperature limit,
where a Yang-Mills theory is asymptotically free.

In the lattice simulations of Refs.\thinspace \cite%
{millerrev,miller97,miller96}, based on the lattice works of Refs. \cite%
{neew}, a striking result was found: the trace of the stress-energy tensor $%
\theta _{\mu \mu }\equiv \rho -3p$ of pure SU(2) and SU(3) Yang-Mills
thermodynamics in the deconfining phase does $not$ vanish (and neither
decreases) but $grows$ linearly with $T$ for $T\gtrsim 2\,T_{c}$ where $%
T_{c} $ denotes the phase boundary. As emphasized in \cite{millerrev}, this
result means that the system is strong interacting even at high $T$ and puts
strong constraints on models aiming at describing these Yang-Mills theories
at finite $T$. Notice that in \cite{fuzzy} a quadratic rise at high $T$
(instead of linear) of $\theta _{\mu \mu }$ is inferred out of lattice data
in the SU(3) case. On one hand this difference points to the difficulty in
extracting high $T$ behavior for thermodynamical quantities, on the other
hand it confirms that nonperturative physics is at work in the high
temperature domain.

While phenomenological models \cite{weise,dirk,levai} predict that, in
accord with asymptotic freedom \cite{NP2004}, the interaction measure $%
\Delta =\frac{\rho -3p}{T^{4}}$ vanishes for $T\rightarrow \infty $, there
is no general agreement on how fast $\Delta $ approaches zero. Namely, as
discussed above, not only $\Delta $ but the quantity $\rho -3p$ itself
should vanish for $T\rightarrow \infty $ if the thermodynamical system is
described by a noninteracting realtivistic gas of gluons. Notice that in the
framework of resummed perturbation theory nonzero contributions to $\theta
_{\mu \mu }$ start at order $g^{4}$ \cite{laine}, where it is shown that $%
\theta _{\mu \mu }$ keeps growing even for high $T.$

Assuming the correctness of the results of \cite{millerrev}, it is therefore
important to study models which $do$ describe the linear growth of $\theta
_{\mu \mu }$ in order to understand the origin of this property. In this
work we show that within the analytical approach to Yang-Mills
thermodynamics of Ref.\thinspace \cite{ralfrev} the linear growth of $\rho
-3p$ is predicted with slopes compatible with lattice results. This property
is a consequence of the presence of a thermal ground state, which takes into
account the contributions of calorons (i.e. instantons at finite $T$).
Intuitively, this thermal ground state, described by an adjoint-scalar field 
$\phi $, represents an $average$ over such topologically non-trivial
configurations. From a phenomenological perspective the ground-state
contribution to $\rho $ behaves like a temperature dependent bag constant,
see for example \cite{weise}. Furthermore, the introduced field $\phi $
generates a Higgs mechanism: (some of) the gauge modes acquire a
temperature-dependent mass, see details in following Section.

The ground-state contribution to the pressure $p_{gs}$ is negative (like the
bag pressure is) and dominates when $T\searrow T_{c}$. Lattice simulations
of $p$ do not agree on this point. Namely, within the differential method,
close to $T_{c}$ negative pressure is predicted \cite{diff} while the
integral method \cite{karsch} is designed to generate positive pressure.
However, in the present report we investigate $\theta _{\mu \mu }$ only for $%
T\gtrsim 1.5\,T_{c}$, thus well above the phase boundary.

In Ref.\thinspace \cite{zw} a linear growth of $\theta _{\mu \mu }$ was
obtained based on the suppression of the infrared modes by invoking a
modified dispersion relation of Gribov type. The latter circumvents the
well-known problems inherent in a perturbative treatment \cite{linde}. In
the alternative approach of \cite{ralfrev,garfield,HH,Hofmann} the stability
of the infrared sector is guaranteed by the above mentioned thermal ground
state, embodied by an adjoint scalar field $\phi $, taking into account
nontrivial-topology fluctuations. Remarkably, this approach predicts the
slopes $\alpha $ in $\theta _{\mu \mu }=\alpha T$ once the Yang-Mills scale $%
\Lambda $ (or alternatively $T_{c}$) is fixed.

The present analysis is not intended to discuss phenomenological fits to
lattice curves (for work in that respect see \cite{weise,levai,dirk} and the
review \cite{dirkrev}). Rather, our goal is to show that the
(asymptotically) linear growth of $\theta _{\mu \mu }$ is related to the
presence of a nontrivial and temperature-dependent ground-state \cite%
{ralfrev,garfield}.

The paper is organized as follows: First we briefly review the approach of
Ref.\thinspace \cite{ralfrev} and list expressions for $\rho $ and $p$ in
the SU(2) case. Subsequently, we show numerically and analytically that
within this approach $\theta _{\mu \mu }=\alpha T$ for $T\gg T_{c}$ and
compare $\alpha $ with the results of \cite{millerrev,miller97,zw}. We then
discuss the SU(3) case and give our conclusions.

\section{\ The SU(2) case}

\emph{\ }In Ref.\thinspace \cite{ralfrev} an adjoint scalar field $\phi $ is
derived which, upon spatial coarse graining, describes the dynamics of
nontrivial, BPS-saturated topological configurations \cite{Nahm,LL} giving
rise to an energy density and pressure of the thermal ground state. The
field $\phi $ acts as a background to the dynamics of topologically \textsl{%
trivial} fluctuations.

The modulus $\left\vert \phi \right\vert =\sqrt{\frac{\Lambda ^{3}}{2\pi \,T}%
}$ exhibits a dependence on $T$ and on $\Lambda $. The corresponding
potential is $V(|\phi |^{2})=\Lambda ^{6}/|\phi |^{2}$ \cite%
{ralfrev,garfield}. In unitary gauge the effective Lagrangian describing a
thermalized, pure SU(2) Yang-Mills system reads: 
\begin{equation}
\mathcal{L}_{{\tiny \mbox{eff}}}^{u.g.}=\mathcal{L}\left[ a_{\mu }\right] =%
\frac{1}{4}\left( G_{E}^{a,\mu \nu }[a_{\mu }]\right) ^{2}+2e^{2}\left\vert
\phi \right\vert ^{2}\left( \left( a_{\mu }^{1}\right) ^{2}+\left( a_{\mu
}^{2}\right) ^{2}\right) +2\frac{\Lambda ^{6}}{\left\vert \phi \right\vert
^{2}}\,,  \label{lagfirstord}
\end{equation}%
where $e=e(T)$ denotes the temperature dependent \textsl{effective} gauge
coupling (see below) which enters both into the effective field strength $%
G_{E}^{a,\mu \nu }$ and into the mass $m$ for the fields $a_{\mu }^{1,2}$.
One has 
\begin{equation}
m^{2}=m(T)^{2}=m_{1}^{2}=m_{2}^{2}=4e^{2}\left\vert \phi \right\vert
^{2}\,,\ \ \ \ m_{3}^{2}=0\,.  \label{masssp}
\end{equation}%
that is a temperature-dependent mass is generated in a nonperturbative
fashion by the adjoint-scalar field $\phi .$ From the effective Lagrangian (%
\ref{lagfirstord}) one derives, on the one-loop level (accurate to about
0.1\% \cite{Hofmann,Schwarz}), the energy density $\rho $ and the pressure $%
p $ as 
\begin{equation}
\rho =\rho _{3}+\rho _{1,2}+\rho _{gs}\,,\ \ \text{ }p=p_{3}+p_{1,2}+p_{gs}%
\,,  \label{rhop}
\end{equation}%
where $\rho _{1,2}$ denotes the sum over the two massive modes. Explicitly
we have: 
\begin{eqnarray}
\rho _{3} &=&2\,\frac{\pi ^{2}}{30}\,T^{4},\text{ }\rho
_{1,2}=6\,\int_{0}^{\infty }\frac{dk\,k^{2}}{2\pi ^{2}}\frac{\sqrt{%
m^{2}+k^{2}}}{\exp (\frac{\sqrt{m^{2}+k^{2}}}{T})-1},\text{ }\rho _{gs}=2%
\frac{\Lambda ^{6}}{\left\vert \phi \right\vert ^{2}}=4\pi \Lambda ^{3}T\,.
\\
p_{3} &=&2\frac{\pi ^{2}}{90}T^{4},\text{ }p_{1,2}=-6\,T\int_{0}^{\infty }%
\frac{dk\,k^{2}}{2\pi ^{2}}\ln \left( 1-e^{-\frac{\sqrt{m^{2}+k^{2}}}{T}%
}\right) \,,\text{ }p_{gs}=-\rho _{gs}\,.
\end{eqnarray}%
Let us now rewrite the system in terms of dimensionless quantities:%
\begin{equation}
\overline{\rho }=\frac{\rho }{T^{4}},\text{ }\ \overline{p}=\frac{p}{T^{4}},%
\text{ }\ \lambda =\frac{2\pi T}{\Lambda },\text{ }\ a(\lambda )=\frac{m(T)}{%
T}=2\frac{e(T)}{T}\left\vert \phi \right\vert \,,
\end{equation}%
where the function $a=a(\lambda )$ is introduced for later use. In
dependence of the dimensionless temperature $\lambda $ the dimensionless
energy density $\bar{\rho}$ and pressure $\bar{p}$ read:%
\begin{equation}
\overline{\rho }_{3}=2\,\frac{\pi ^{2}}{30}\,,\text{ }\overline{\rho }_{1,2}=%
\frac{3}{\pi ^{2}}\int_{0}^{\infty }dx\,\frac{x^{2}\sqrt{x^{2}+a^{2}}}{e^{%
\sqrt{x^{2}+a^{2}}}-1}\,,\text{ }\overline{\rho }_{gs}=\frac{2(2\pi )^{4}}{%
\lambda ^{3}}\,,  \label{rhoad}
\end{equation}%
\begin{equation}
\overline{p}_{3}=2\,\frac{\pi ^{2}}{90}\,,\ \text{ }\overline{p}_{1,2}=-%
\frac{3}{\pi ^{2}}\int_{0}^{\infty }dx\,x^{2}\ln \left( 1-e^{-\sqrt{%
x^{2}+a^{2}}}\right) \,,\ \text{ }\overline{p}_{gs}=-\text{ }\overline{\rho }%
_{gs}\,.  \label{pad}
\end{equation}%
By virtue of the Legendre transformation 
\begin{equation}
\rho =T\frac{dP}{dT}-P\iff \overline{\rho }=\lambda \frac{d\overline{p}}{%
d\lambda }+3\overline{p}  \label{tsc}
\end{equation}%
the evolution equation for $a=a(\lambda )$ follows:%
\begin{eqnarray}  \label{ev.e}
1 &=&-\frac{6\lambda ^{3}}{(2\pi )^{6}}\left( \lambda \frac{da}{d\lambda }%
+a\right) aD(a),\text{ }  \label{aeq} \\
D(a) &=&\int_{0}^{\infty }dx\frac{x^{2}}{\sqrt{x^{2}+a^{2}}}\frac{1}{e^{%
\sqrt{x^{2}+a^{2}}}-1},\text{ }a(\lambda _{in})\ll 1.
\end{eqnarray}%
There exits a low-temperature attractor to the evolution described by
Eq.\thinspace (\ref{ev.e}) with a logarithmic pole at the critical
temperature $\lambda _{c}=13.89$ \cite{ralfrev}. At $\lambda _{c}$ the
fluctuations $a_{\mu }^{1,2}$ decouple thermodynamically. The effective
coupling is given as $e=e(\lambda )=a(\lambda )\lambda ^{3/2}/4\pi $. For $%
a\ll 1$ the coupling $e$ is constant: $e=\sqrt{8}\pi $. In fact 
\begin{equation}
a(\lambda )=\frac{8\sqrt{2}\pi ^{2}}{\lambda ^{3/2}}  \label{alamdalim}
\end{equation}%
then is a solution to Eq.\thinspace (\ref{ev.e}). This plateau is reached
rapidly for increasing $\lambda $, see figures in \cite{ralfrev}.

\section{\emph{\ }Linear growth of $\protect\theta _{\protect\mu \protect\mu %
}$}

We now turn to the linear rise of $\theta _{\mu \mu }$ within the approach
described in the previous Section. In therms of dimensionless quantities one
has 
\begin{equation}
\frac{\theta _{\mu \mu }}{\Lambda ^{4}}=\frac{\rho -3p}{\Lambda ^{4}}=\left( 
\frac{\lambda }{2\pi }\right) ^{4}(\overline{\rho }-3\overline{p})\,.
\end{equation}%
A plot of the quantity $\frac{\theta _{\mu \mu }}{\Lambda ^{4}}$ is shown in
Fig.\thinspace 2. Notice the linear growth.

\begin{figure}[tbp]
\begin{center}
\leavevmode
\leavevmode
\vspace{4.5cm} \includegraphics{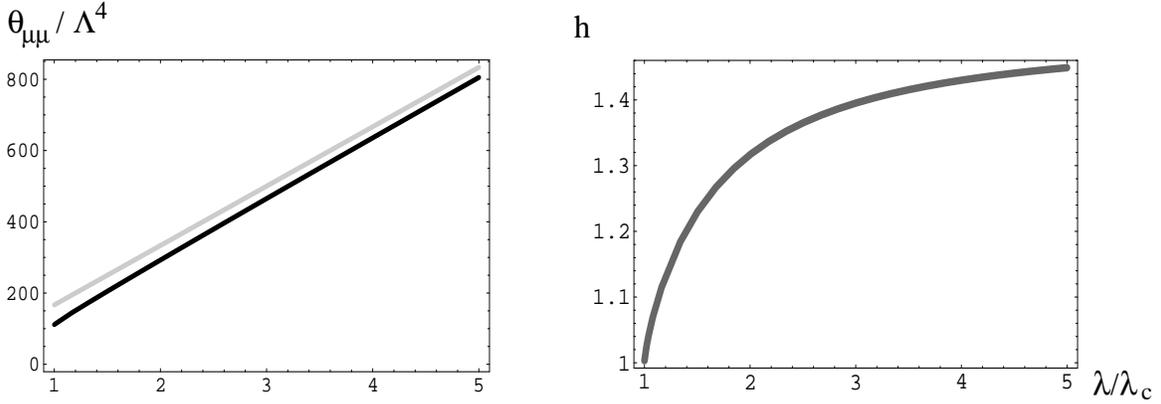}
\end{center}
\caption{The quantities $\frac{\protect\theta _{\protect\mu \protect\mu }}{%
\Lambda ^{4}}$ (left panel, gray curve: large-$T$ behavior; black curve:
finite-$T$ behavior) and $h$ (right panel) as functions of $\frac{\protect%
\lambda }{\protect\lambda _{c}}=\frac{T}{T_{c}}$ in the case of SU(2).}
\label{Fig-1}
\end{figure}
Asymptotically ($\lambda \gg \lambda _{c}$) this linear rise can be shown to
hold analytically. To this end let us consider the function%
\begin{equation}
h(\lambda )=\frac{\rho -3p}{4\rho _{gs}}=\frac{\overline{\rho }-3\overline{p}%
}{4\overline{\rho }_{gs}}=1+\frac{\overline{\rho }_{1,2}-3\overline{p}_{1,2}%
}{4\overline{\rho }_{gs}}\,.  \label{hdef}
\end{equation}%
For $a\ll 1$ (i.e. $\lambda \gg \lambda _{c}$) we have \cite{DolanJackiw}: 
\begin{equation}
\overline{\rho }_{1,2}=6\,\frac{\pi ^{2}}{30}-\frac{1}{4}a^{2}+\cdots ,\text{
}\overline{p}_{1,2}=6\,\frac{\pi ^{2}}{90}-\frac{1}{4}a^{2}+\cdots \,.
\end{equation}%
As a consequence, the function $h=h(\lambda )$ becomes:%
\begin{equation}
h(\lambda )=1+\frac{\frac{1}{2}\,a^{2}}{4\frac{2(2\pi )^{4}}{\lambda ^{3}}}%
+\cdots =1+\frac{\lambda ^{3}a^{2}}{16(2\pi )^{4}}+\cdots =\frac{3}{2}%
+\cdots \,,
\end{equation}%
where Eq.\thinspace (\ref{alamdalim}) has been used. Thus $h(\lambda )$ is
approximately constant for $a\ll 1$. The numerical behavior for the function 
$h=h(\lambda )$ is plotted in the right panel of Fig.\thinspace \ref{Fig-1}.
Notice that the asymptotic value $h(\lambda )\equiv \frac{3}{2}$ is
practically reached for $\lambda \geq 5\,\lambda _{c}$.

By virtue of Eq.\,(\ref{hdef}) we have asymptotically (corresponding to the
gray curve in the left panel of Fig\,\ref{Fig-1}):%
\begin{equation}
\rho-3p=6\rho _{gs}=24\pi \Lambda ^{3}T\,.
\end{equation}
It is interesting that $\theta _{\mu \mu }$ splits as $\theta _{\mu \mu
}=6\rho _{gs}=4\rho_{gs}+2\rho _{gs}$ where the first summand is the direct
contribution of the ground state while the second summand arises from the
massive modes $a_{\mu }^{1,2}$. That is, the mass $m=m(T)$ behaves in such a
way that fluctuations generate a linear contribution to $\theta _{\mu \mu }$
at high $T$.

Let us now compare our prediction for the slope with lattice results. To
this end we relate the Yang-Mills scale $\Lambda $ to the critical
temperature $T_{c}$ as%
\begin{equation}
\Lambda =\frac{2\pi T_{c}}{\lambda _{c}}\simeq 0.45\,T_{c}\,,  \label{slope}
\end{equation}%
where $\lambda _{c}=13.89$ has been used. Then we have asymptotically 
\begin{equation}
\theta _{\mu \mu }=\rho -3p=\frac{192\,\pi ^{4}}{\lambda _{c}^{3}}%
T_{c}^{3}T\simeq 7\,T_{c}^{3}T\,.  \label{slopeff}
\end{equation}%
In the work of \cite{millerrev} a value of $T_{c}=0.290$\thinspace GeV was
used. Thus Eq.\thinspace (\ref{slopeff}) asymptotically predicts $\theta
_{\mu \mu }\simeq 1.7\,T\,\mbox{GeV}^{3}$. Notice that according to our
approach the slope increases from $\simeq 1.4$\thinspace GeV$^{3}$ to the
asymptotic value $1.7$\thinspace GeV$^{3}$ for $1.4$\thinspace $T_{c}\leq
T\leq 5\,T_{c}$. Thus our prediction for the slope is in agreement with that
of Miller \cite{millerrev} whose lattice simulation yields $\sim 1.5$%
\thinspace GeV$^{3}$ (read off from his Fig.\thinspace 1). In Ref.\thinspace 
\cite{zw} a slope of $\sim 0.2$\thinspace GeV$^{3}$ was obtained for the
SU(2) case which also is in qualitative agreement with our result. Notice
that the lattice simulations of \cite{langfeld} observes a linear behavior
of $\theta _{\mu \mu }$, too.

\begin{figure}[tbp]
\begin{center}
\leavevmode
\leavevmode
\vspace{5.5cm} \includegraphics{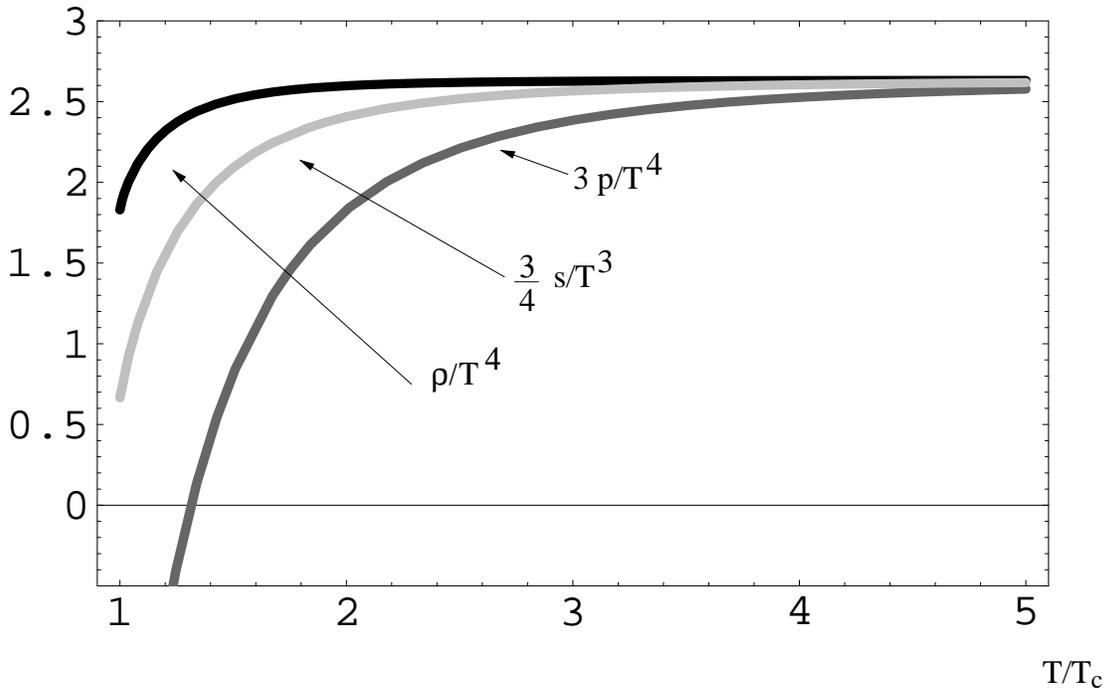}
\end{center}
\caption{Scaled energy density $\protect\rho$ (black), pressure $p$ (dark
gray), and entropy density $s$ (light gray) in the deconfining phase of
SU(2) Yang-Mills thermodynamics.}
\label{Fig-2}
\end{figure}

In Fig. 2 we report the (scaled) energy density $\rho $, pressure $p$, and
entropy density $s$ as functions of temperature. A detailed comparison of
these results with those obtained on the lattice (for both differential and
integral method and for the SU(3) case) is carried out in \cite{ralfrev}.
While there is good agreement for the entropy density $s$ (for which the
ground state is not relevant) the pressure $p$ becomes negative close to the
phase boundary in virtue of the negative contribution of the ground state.
While this is in qualitative agreement with the result of the differential
method, it is not with that of the integral method. This fact, as emphaiszed
in the Introduction, represents a motivation to perform the present work
where lattice results safely above the phase boundary, such as the behavior
of $\theta _{\mu \mu }$ at high $T$, are compared with the effective
approach under study.

\section{The SU(3) case}

In the SU(3) case the behavior of $\theta _{\mu \mu }$ is qualitatively
similar to the SU(2) case. We only report on some relevant formulas and
briefly discuss their consequences. The modulus of the scalar field $\phi $
is exactly the same. As shown in \cite{ralfrev} out of the eight
coarse-grained gauge modes four acquire a mass $m_{1}=e\left\vert \phi
\right\vert $ (contributing to $\rho $ and $p$ by $\rho _{1}$ and $p_{1}$),
two a mass $m_{2}=2e\left\vert \phi \right\vert $ ($\rho _{2}$ and $p_{2}$),
and two stay massless ($\rho _{3}$ and $p_{3}$). Explicitly, we have ($%
a=m_{1}/T$): 
\begin{eqnarray}
\overline{\rho }_{3} &=&4\,\frac{\pi ^{2}}{30}\,,\ \text{ }\overline{\rho }%
_{1}=\frac{6}{\pi ^{2}}\int_{0}^{\infty }dx\,\frac{x^{2}\sqrt{x^{2}+a^{2}}}{%
e^{\sqrt{x^{2}+a^{2}}}-1}\,, \\
\overline{\rho }_{2} &=&\frac{3}{\pi ^{2}}\int_{0}^{\infty }dx\,\frac{x^{2}%
\sqrt{x^{2}+(2a)^{2}}}{e^{\sqrt{x^{2}+(2a)^{2}}}-1}\,,\ \text{ }\overline{%
\rho }_{gs}=\frac{2(2\pi )^{4}}{\lambda ^{3}}\,,
\end{eqnarray}%
\begin{eqnarray}
\overline{p}_{3} &=&4\,\frac{\pi ^{2}}{90}\,,\ \text{ }\overline{p}_{1}=-%
\frac{6}{\pi ^{2}}\int_{0}^{\infty }dx\,x^{2}\ln \left( 1-e^{-\sqrt{%
x^{2}+a^{2}}}\right) \,,\text{ } \\
\overline{p}_{2} &=&-\frac{3}{\pi ^{2}}\int_{0}^{\infty }dx\,x^{2}\ln \left(
1-e^{-\sqrt{x^{2}+(2a)^{2}}}\right) \,,\ \text{ }\overline{p}_{gs}=-\text{ }%
\overline{\rho }_{gs}\,.
\end{eqnarray}%
Eq.\thinspace (\ref{tsc}) assumes the following form:%
\begin{equation}
1=-\frac{12\lambda ^{3}}{(2\pi )^{6}}\left( \lambda \frac{da}{d\lambda }%
+a\right) \left( aD(a)+2aD(2a)\right) \,.  \label{e.ev3}
\end{equation}%
The asymptotic solution to Eq.\thinspace (\ref{e.ev3}) reads $a(\lambda )=%
\frac{8}{\sqrt{3}}\pi ^{2}\lambda ^{-3/2}$, and the effective coupling
reaches a plateau value of $e=\frac{4}{\sqrt{3}}\pi .$ The value for $%
\lambda _{c}$ is $\lambda _{c}=9.475$ \cite{ralfrev} where the coupling and
all masses diverge. The function $h(\lambda )$, defined in (\ref{hdef}), has
an asymptotic value of $3/2$ like in the SU(2) case. Thus the asymptotically
linear behavior $\theta _{\mu \mu }=\rho -3p=24\pi \Lambda ^{3}T$ holds.
Relating $\Lambda $ to the critical temperature $T_{c}$ as in Eq. (\ref%
{slope}), we take into account the SU(3)-value $\lambda _{c}=9.475$. This
yields:%
\begin{equation}
\theta _{\mu \mu }=\rho -3p=\frac{192\pi ^{4}}{\lambda _{c}^{3}}%
T_{c}^{3}T\simeq 21.9\,T_{c}^{3}T\,.
\end{equation}%
As in Ref.\thinspace \cite{millerrev} we set $T_{c}=0.264$\thinspace GeV and
thus find $\theta _{\mu \mu }\simeq 0.4\,T\,\mbox{GeV}^{3}$ for $T\gg T_{c}$%
. In \cite{millerrev} a somewhat smaller slope of $\simeq 0.2$\thinspace GeV$%
^{3}$ is extracted. In \cite{zw} a slope of $0.63$\thinspace GeV$^{3}$ was
estimated.

\section{Conclusions}

In this paper we studied the high-temperature behavior of the trace of the
stress-energy tensor $\theta _{\mu \mu }=\rho -3p.$ This quantity measures
the breaking of scale invariance and vanishes for a relativist gas of free
gluons: recent lattice works \cite{millerrev,miller97} show a linear rise
for high $T$ (see however also \cite{fuzzy}, where the rise is quadratic).
This means that strong interactions still play a decisive role even in the
high temperature domain.

Theoretical studies should therefore be concerned with the predicted
behavior of $\theta _{\mu \mu }$ at high temperatures. We have shown that
the linear growth of $\theta _{\mu \mu }$ for high $T$ is an analytic
prediction of the approach to Yang-Mills thermodynamics \cite{ralfrev},
which involves a thermal ground state emerging as an average over instantons
at finite $T$. That is, even in the high-$T$ limit the dynamic of these
field configurations is important and can be detected by lattice
simulations. In both cases, SU(2) and SU(3), we evaluated the slope of $%
\theta _{\mu \mu }$ and the results $\theta _{\mu \mu }\simeq
7\,T_{c}^{3}T\, $and $\theta _{\mu \mu }\simeq 21.9\,T_{c}^{3}T\,,\ $where $%
T_{c}$ is the deconfining temperature, are found for high $T$, respectively.
We compared to the lattice results of Miller \cite{millerrev,miller97} and
to the theoretical work of Zwanziger \cite{zw} finding good agreement. In
the latter a momentum-dependent, universal modification of the dispersion
relation for propagating, fundamental gluon fields, motivated by the
reduction of the physical state space a la Gribov, is introduced. Our
approach and \cite{zw} share the property of a nonpertrubative gluon mass
leading to the linear growth of $\theta _{\mu \mu }$. However, while in our
case the mass is a temperature-dependent coarse grained quantity and the
ground state itsself contributes to $\theta _{\mu \mu },$ this is not the
case in \cite{zw}. A more detailed comparison would be surely interesting
but goes beyond the scope of the present work.

We believe that future studies on the high $T$ behavior of $\theta _{\mu \mu
}$ are therefore very important, both numerically and analytically, in order
to confirm the linear growth and to relate it to nonperturbative aspects of
Yang-Mills theories.


\bigskip

\begin{thebibliography}{99}
\bibitem{millerrev} D.~E.~Miller, 
arXiv:hep-ph/0608234. 

\bibitem{miller97} D.~E.~Miller, 
Acta Phys.\ Polon.\ B \textbf{28} (1997) 2937. 

\bibitem{zw} D.~Zwanziger, 
Phys.\ Rev.\ Lett.\ \textbf{94} (2005) 182301. 

\bibitem{miller96} G.~Boyd and D.~E.~Miller, 
arXiv:hep-ph/9608482. 

\bibitem{neew} J.~Engels, F.~Karsch and K.~Redlich, 
Nucl.\ Phys.\ B \textbf{435} (1995) 295 [arXiv:hep-lat/9408009]. 
G.~Boyd, J.~Engels, F.~Karsch, E.~Laermann, C.~Legeland, M.~Lutgemeier and
B.~Petersson, 
Nucl.\ Phys.\ B \textbf{469} (1996) 419 [arXiv:hep-lat/9602007]. 

\bibitem{fuzzy} R.~D.~Pisarski, 
arXiv:hep-ph/0612191. 

\bibitem{weise} R.~A.~Schneider and W.~Weise, 
Phys.\ Rev.\ C \textbf{64} (2001) 055201. 

\bibitem{dirk} D.~H.~Rischke, J.~Schaffner, M.~I.~Gorenstein, A.~Schaefer,
H.~Stoecker and W.~Greiner, 
Z.\ Phys.\ C \textbf{56} (1992) 325. 

\bibitem{levai} P.~Levai and U.~W.~Heinz, 
Phys.\ Rev.\ C \textbf{57} (1998) 1879. 

\bibitem{NP2004} D. J. Gross and Frank Wilczek, Phys. Rev. D \textbf{8},
3633 (1973).\newline
D. J. Gross and Frank Wilczek, Phys. Rev. Lett. \textbf{30}, 1343 (1973).%
\newline
H. David Politzer, Phys. Rev. Lett. \textbf{30}, 1346 (1973).\newline
H. David Politzer, Phys. Rept. \textbf{14}, 129 (1974). 

\bibitem{laine} M.~Laine and Y.~Schroder, 
Phys.\ Rev.\ D \textbf{73} (2006) 085009 [arXiv:hep-ph/0603048]. 
K.~Kajantie, M.~Laine, K.~Rummukainen and Y.~Schroder, 
Phys.\ Rev.\ D \textbf{67} (2003) 105008 [arXiv:hep-ph/0211321]. 

\bibitem{ralfrev} R.~Hofmann, 
Int.\ J.\ Mod.\ Phys.\ A\textbf{20} (2005) 4123, Erratum-ibid.\ A \textbf{21}
(2006) 6515.\newline
R.~Hofmann, Mod. Phys. Lett. A\textbf{21}, 999 (2006), Erratum-ibid. A 
\textbf{21}, 3049 (2006). 

\bibitem{diff} F.~R.~Brown, N.~H.~Christ, Y.~F.~Deng, M.~S.~Gao and
T.~J.~Woch, 
Phys.\ Rev.\ Lett.\ \textbf{61} (1988) 2058. 

\bibitem{karsch} J.~Engels, J.~Fingberg, F.~Karsch, D.~Miller and M.~Weber, 
Phys.\ Lett.\ B \textbf{252} (1990) 625. 

\bibitem{linde} A.~D.~Linde, 
Phys.\ Lett.\ B \textbf{96} (1980) 289. 

\bibitem{garfield} F.~Giacosa and R.~Hofmann, 
arXiv:hep-th/0609172. 

\bibitem{HH} U. Herbst and R. Hofmann, arXiv:hep-th/0411214.

\bibitem{Hofmann} R. Hofmann, arXiv:hep-th/0609033.

\bibitem{dirkrev} D.~H.~Rischke, 
Prog.\ Part.\ Nucl.\ Phys.\ \textbf{52} (2004) 197. 

\bibitem{Nahm} W. Nahm, Lect. Notes in Physics. 201, eds. G. Denaro, e.a.
(1984) p. 189.

\bibitem{LL} K.-M. Lee and C.-H. Lu, Phys. Rev. D \textbf{58}, 025011 (1998).%
\newline
T. C. Kraan and P. van Baal, Nucl. Phys. B \textbf{533}, 627 (1998).\newline
T. C. Kraan and P. van Baal, Phys. Lett. B \textbf{435}, 389 (1998).

\bibitem{Schwarz} M.~Schwarz, R.~Hofmann and F.~Giacosa, 
arXiv:hep-th/0603078 (in press Int. J. Mod. Phys. A). 

\bibitem{DolanJackiw} L. Dolan and R. Jackiw, Phys. Rev. D \textbf{9}, 3320
(1974).


\bibitem{langfeld} K.~Langfeld, E.~M.~Ilgenfritz, H.~Reinhardt and G.~Shin, 
Nucl.\ Phys.\ Proc.\ Suppl.\ \textbf{106} (2002) 501
[arXiv:hep-lat/0110024]. 
\end{thebibliography}
\end{document}